\documentclass[sigconf]{acmart}
\AtBeginDocument{%
  }
\usepackage{subfig}
\usepackage{multirow}
\usepackage{url}
\usepackage{hyperref}

\copyrightyear{2026}
\acmYear{2026}
\setcopyright{cc}
\setcctype{by}
\acmConference[WWW Companion '26]{Companion Proceedings of the ACM Web Conference 2026}{April 13--17, 2026}{Dubai, United Arab Emirates}
\acmBooktitle{Companion Proceedings of the ACM Web Conference 2026 (WWW Companion '26), April 13--17, 2026, Dubai, United Arab Emirates}
\acmPrice{}
\acmDOI{10.1145/3774905.3793103}
\acmISBN{979-8-4007-2308-7/2026/04}

\begin{document}

\title{STIndex: A Context-Aware Multi-Dimensional Spatiotemporal Information Extraction System}

\author{Wenxiao Zhang$^{\dagger, *}$}
\email{wenxiao.zhang@research.uwa.edu.au}
\affiliation{%
  \institution{The University of Western Australia}
  \city{Perth}
  \country{Australia}
}
\author{Yu Liu$^{\dagger,*}$}
\email{liuyu@iie.ac.cn}
\affiliation{%
  \institution{Institute of Information Engineering, Chinese Academy of Sciences}
  \institution{University of Chinese Academy of Sciences, Beijing, China}
  \city{}
  \country{}
}

\author{Qiang Sun}
\email{pascal.sun@research.uwa.edu.au}
\affiliation{%
  \institution{The University of Western Australia}
  \city{Perth}
  \country{Australia}
}

\author{Yihao Ding}
\email{yihao.ding@uwa.edu.au}
\affiliation{%
  \institution{The University of Western Australia}
  \city{Perth}
  \country{Australia}
}

\author{Sirui Li}
\email{sirui.li@murdoch.edu.au}
\affiliation{%
  \institution{Murdoch University}
  \city{Perth}
  \country{Australia}
}

\author{Yanbing Liu}
\email{liuyanbing@iie.ac.cn}
\affiliation{%
  \institution{Institute of Information Engineering, Chinese Academy of Sciences}
  \institution{University of Chinese Academy of Sciences, Beijing, China}
  \city{}
  \country{}
}

\author{Jin B. Hong$^{\dagger}$}
\email{jin.hong@uwa.edu.au}
\affiliation{%
  \institution{The University of Western Australia}
  \city{Perth}
  \country{Australia}
}

\author{Wei Liu$^{\dagger}$}
\email{wei.liu@uwa.edu.au}
\affiliation{%
  \institution{The University of Western Australia}
  \city{Perth}
  \country{Australia}
}

% \author{Wei Liu$^{\dagger}$}
% \email{wei.liu@uwa.edu.au}
% \affiliation{%
%   \institution{The University of Western Australia}
%   \city{Perth}
%   \state{WA}
%   \country{Australia}
% }

% \author{Wenxiao Zhang}
% \email{wenxiao.zhang@research.uwa.edu.au}
% \affiliation{%
%   \institution{The University of Western Australia}
%   \city{Perth}
%   \state{WA}
%   \country{Australia}
% }

\renewcommand{\shortauthors}{Wenxiao Zhang et al.}
%% No italics, no superscripts, not anonymous
%% Use footnote or author note to identify equal contribution and/or contact author info


\begin{abstract}
Extracting structured knowledge from unstructured data still faces practical limitations: entity and event extraction pipelines remain brittle, knowledge graph construction requires costly ontology engineering, and cross-domain generalization is rarely production-ready. In contrast, space and time provide universal contextual anchors that naturally align heterogeneous information and benefit downstream tasks such as retrieval and reasoning. We introduce \textbf{STIndex}, an end-to-end system that structures unstructured content into a multidimensional spatiotemporal data warehouse. Users define domain-specific analysis dimensions with configurable hierarchies, while large language models perform context-aware extraction and grounding. \textbf{STIndex} integrates document-level memory, geocoding correction, and quality validation, and offers an interactive analytics dashboard for visualization, clustering, burst detection, and entity network analysis. In evaluation on a public health benchmark, \textbf{STIndex} improves spatiotemporal entity extraction F1 by 4.37\% (GPT-4o-mini) and 3.60\% (Qwen3-8B). A live demonstration and open-source code are available at \url{https://stindex.ai4wa.com/dashboard}.

% % 第一句是不是可以说entity extration, event extration, knowledge graph construction still blocked due to the xxx, blocking from production application, however, spatiotemporal information is unified across domains, and are very important for information retrieval, reasoning and benefit downstream tasks like RAG, QA, etc.
% Extracting structured spatiotemporal information from unstructured web content remains challenging due to entity ambiguity, geocoding errors, and quality control issues. We present \textbf{STIndex}, an end-to-end system that combines context-aware LLM extraction and interactive visualization for multi-dimensional spatiotemporal analysis. Key features include: \textbf{(1)} configurable custom dimensions enabling domain-agnostic spatiotemporal extraction, 
% % 这里感觉有点交代不清楚，custom dimension 是啥意思？spatiotempoal enabled custom dimensions?
% \textbf{(2)} a context-aware extraction pipeline with document-level memory and post-processing tools, and \textbf{(3)} an out-of-box interactive dashboard with spatiotemporal clustering, burst detection, and entity network analysis. We demonstrate \textbf{STIndex} with a public health surveillance case study on real-world health event reports. Evaluation of 500 documents shows that context-awareness improves combined F1 by 4.37\% for GPT-4o-mini and 3.60\% for Qwen3-8B. 
% % 这里的evaluation的任务是什么？
% Project homepage at \url{https://stindex.ai4wa.com/}, a demonstration interactive dashboard at \url{https://stindex.ai4wa.com/dashboard}. 
% Code is open-source at GitHub: \url{https://github.com/MoeBuTa/STIndex}.

\noindent\rule{1.7cm}{0.4pt}\\
{\footnotesize $^{*}$Equal contribution.}\\
{\footnotesize $^{\dagger}$Corresponding author.}
\end{abstract}





\maketitle
\begin{figure*}[t]
  \centering
  \includegraphics[width=\textwidth]{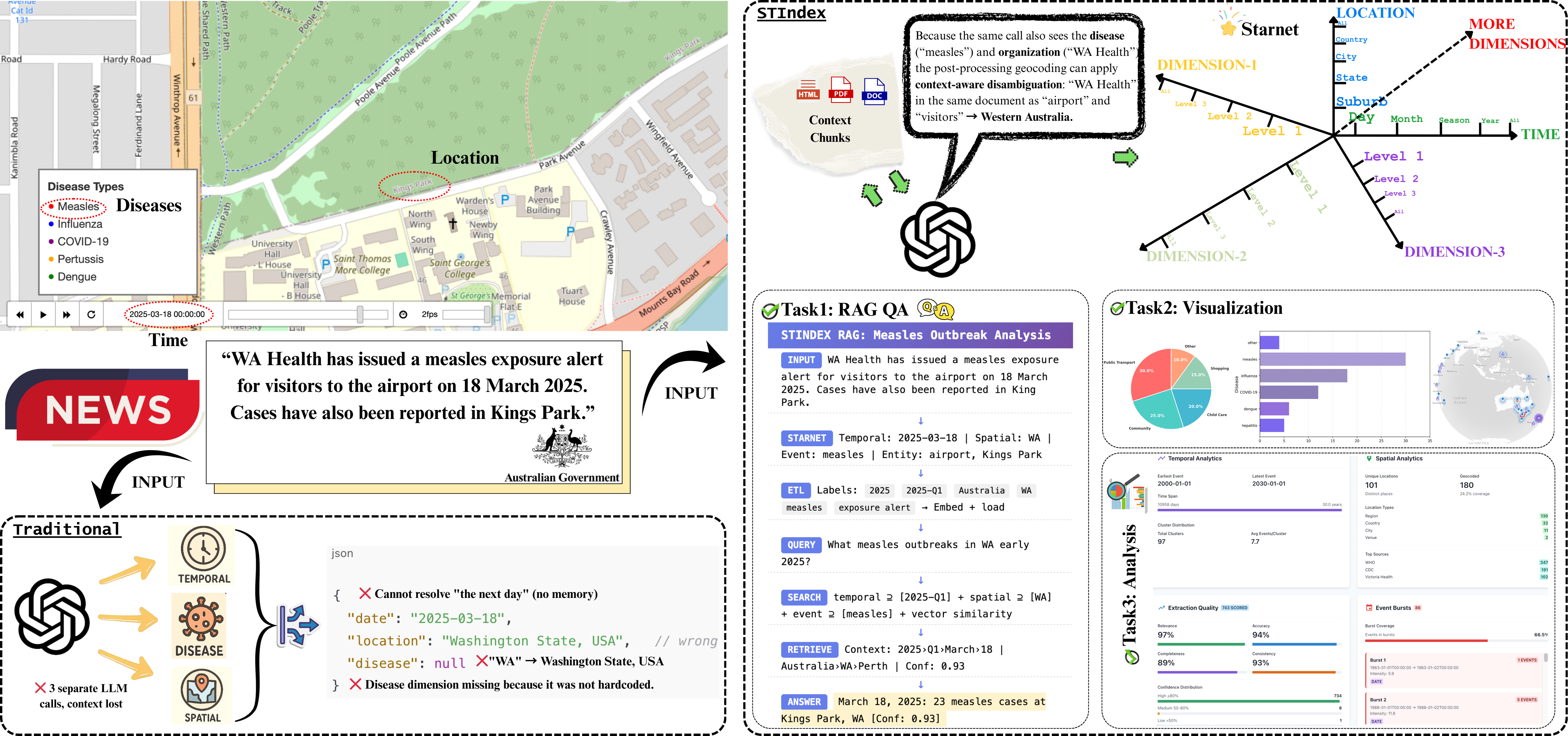}
  % \caption{Motivation example for STIndex. A real public-health alert ("WA Health has issued a measles exposure alert…") contains temporal, spatial, disease, event-type and venue-type signals. \textbf{Left (Traditional):} the pipeline calls LLMs multiple times for temporal/spatial/disease, so there is no shared context; "WA" is resolved to \emph{Washington State, USA} and the disease dimension is missing because it was not hardcoded. \textbf{Right (STIndex):} a single custom-driven extraction call sees all dimensions at once; post-processing can apply context-aware geocoding ("WA Health" + airport + visitors) and multi-level fallback to confirm \emph{Western Australia}, while also filling disease, event type, and venue type.}
  \caption{Public health alert example: a split pipeline loses context and misreads “WA” as Washington, while STIndex’s unified, context-aware extraction correctly resolves it as Western Australia.}
  \Description{A comparison diagram showing two approaches to spatiotemporal information extraction from a public health alert. The left side shows a traditional pipeline with separate extraction calls losing context, incorrectly geocoding WA to Washington State. The right side shows STIndex's unified approach correctly identifying Western Australia through context-aware processing.}
  \label{fig:stindex-motivation}
\end{figure*}

\section{Introduction}

Unstructured data continues to grow rapidly across domains, yet real-world adoption of structured extraction remains limited. Despite progress in entity and relation extraction, event detection, and knowledge graph construction~\cite{iepile2024}, existing pipelines still struggle with ambiguity, domain transfer, and production-level robustness~\cite{chen2022frontiers}. In contrast, spatiotemporal information is naturally shared across domains and offers a stable organizational context for unstructured data. Motivated by the success of data warehouses, we propose to treat space and time as universal anchoring dimensions and define domain-specific analysis dimensions as configurable hierarchies. Instead of requiring complex ontologies, this multidimensional and hierarchical design enables scalable cross-domain adoption and supports downstream tasks such as retrieval-augmented generation~(RAG), visualization, analysis, and reasoning.

% Extracting spatiotemporal information from unstructured text is fundamental for various domains such as public health surveillance, disaster response, and news analysis. Recent advances in temporal extraction, spatial geocoding, and general information extraction demonstrate strong performance on specialized tasks, with systems achieving high accuracy on temporal relations~\cite{typedmarkers2023}, significant improvements in geocoding precision~\cite{llmgeoparsing2024}, and comprehensive schema-based extraction pipelines~\cite{iepile2024,llmaix2025}.
% % spatiotempral dependency, they normally exists together for meaningful information, should not be fragmented

However, existing systems to extract spatial and temporal information remain fragmented: temporal~\cite{heideltime2013,typedmarkers2023} and spatial~\cite{mordecai2017,llmgeoparsing2024} extractors usually operate independently, creating pipeline composition errors, context loss, and lacking unified frameworks for downstream analytics. Existing approaches face three critical gaps. First, specialized extractors lack preprocessing for diverse inputs (PDF, HTML, DOCX) and fail to provide complete pipelines from source to insight. Second, architectural fragmentation causes temporal and spatial systems to work separately, losing the inter-dimensional context needed for disambiguation across document chunks. Third, quality control challenges arise from noisy LLM outputs with false positives requiring manual verification.

We present \textbf{STIndex}, an end-to-end system that enables users to define domain-specific multidimensional schemas—composed of spatial, temporal, and additional semantic dimensions—and leverage them for context-aware information extraction and interactive analytics. 
Our key contributions are: \textbf{(1)} a domain-agnostic extraction framework that supports configurable multidimensional schemas without code modifications; \textbf{(2)} unified single-call LLM extraction with document-level memory to resolve ambiguous references, enhanced by specialized post-processing modules; and \textbf{(3)} an out-of-the-box analytics interface providing spatiotemporal visualization together with clustering, burst detection, and graph-based relation analysis.  STIndex is publicly available on PyPI and can be installed via \texttt{pip install stindex}, enabling immediate deployment for researchers and practitioners.

\section{Related Work}
\begin{figure*}[t]
  \centering
  \includegraphics[width=\textwidth]{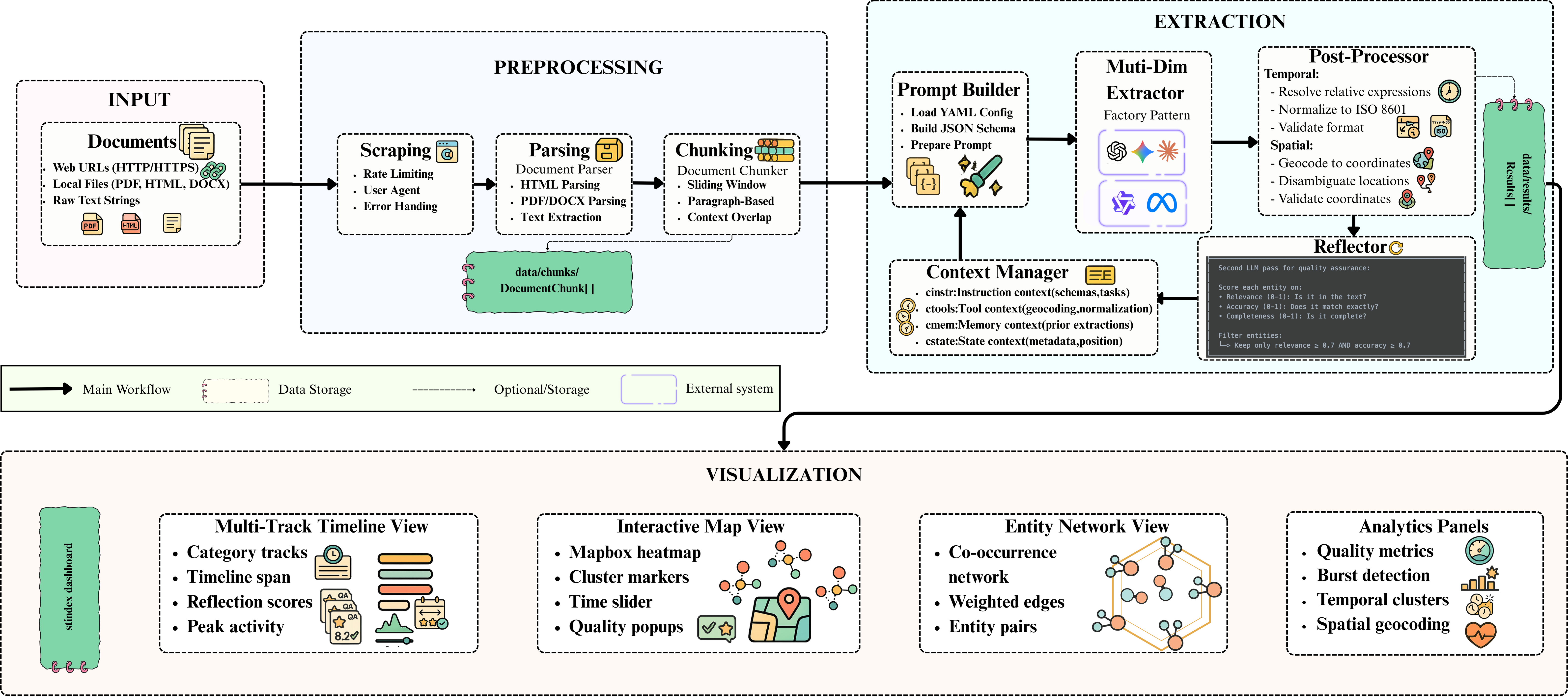}
  \caption{The Overview of the STIndex System Architecture}
  \label{fig:stindex-overview}
\end{figure*}

% male, is wrong,. left side the wrong part, too many x, and right side,  RAG QA  example is not a QA, maybe change the name? and say enable QA or RAG, and then starnet is not correct.

\noindent\textbf{\textit{Spatiotemporal Information Extraction.}} Spatiotemporal information extraction traditionally separates temporal and spatial processing. Temporal systems like HeidelTime~\cite{heideltime2013} provide rule-based normalization with high precision but limited recall, while neural approaches achieve 83.5\% F1 on clinical temporal relations~\cite{typedmarkers2023} using typed markers with BERT. Spatial extraction evolved from geoparsing like Mordecai~\cite{mordecai2017} to LLM-based approaches fine-tuning Mistral and Llama2 with LoRA, achieving 91\% Accuracy@161km~\cite{llmgeoparsing2024}, which is 17\% better than prior methods. These systems operate independently, losing inter-dimensional context for disambiguation.

\noindent\textbf{\textit{End-to-End Information Extraction.}} End-to-end systems demonstrate comprehensive pipeline integration. IEPile~\cite{iepile2024} created the largest schema-based IE corpus with 0.32B tokens across 33 datasets using instruction tuning. LLM-AIx~\cite{llmaix2025} provides complete medical document processing with 87--92\% accuracy, incorporating OCR, parsing, and web interfaces. While these systems address preprocessing and usability, neither specifically targets spatiotemporal information nor provides visualizations revealing temporal-spatial patterns. Domain-specific systems remain fragmented, requiring users to chain separate tools for their specific use cases.

% While these systems address preprocessing and usability, neither specifically targets spatiotemporal information nor provides animated visualizations revealing temporal-spatial patterns. Domain-specific systems remain fragmented, requiring users to chain separate tools for web scraping, document parsing, temporal extraction, spatial geocoding, and visualization.

\section{System Architecture}

\textbf{STIndex.} employs a three-stage architecture with three modules: (1) unstructured data preprocessing, (2) spatiotemporal-aware information extraction, and (3) downstream analysis and visualization. The overall workflow is illustrated in Figure~\ref{fig:stindex-overview}.

\subsection{Unstructured Data Preprocessing Module}

The \textit{PREPROCESSING} module provides unified document ingestion from heterogeneous sources, transforming diverse formats into structured text suitable for extraction. Three input types are supported: web URLs with rate-limited scraping, local files (HTML, PDF, DOCX, TXT) via the unstructured library, and raw text. Document metadata (publication date, source location) is preserved for context-aware extraction. Four chunking strategies, sliding window, paragraph-based, element-based, and semantic chunking, are supported to process long documents, with default configuration using 2000, character chunks with 200, character overlap.

\subsection{Context-Aware Extraction} 
% this requires a better and more proper name

The \textit{EXTRACTION} module implements unified multi-dimensional extraction with context awareness across document chunks, resolving relative temporal expressions ("the next day") and ambiguous spatial references ("the city") through extraction memory.

\noindent\textbf{\textit{Context Engineering.}} The system uses four types of context. \textit{Memory context ({Cmem})} keeps track of previously mentioned entities to resolve relative references. \textit{State context (Cstate)} holds document metadata, and 
\textit{instruction context (Cinstr)} makes sure extractions stay consistent with what came before. \textit{Tool context (Ctools)} allows post-processing tools to access the complete document context.

\noindent\textbf{\textit{Unified Extraction.}} A single LLM call processes all dimensions simultaneously, avoiding context loss and enabling cross-dimensional disambiguation, where custom dimensions like venues constrain spatial extraction, and temporal context resolves ambiguities. 
Four extraction types: normalized extraction converts temporal expressions to ISO 8601, geocoded extraction resolves coordinates using multi-level fallback, categorical extraction maps to controlled vocabularies, and structured extraction handles multi-attribute entities.

\noindent\textbf{\textit{LLM Backend Flexibility.}} The system supports multiple LLM backends: propriety  APIs for state-of-the-art accuracy with per-token costs, and open-source models for self-hosted deployment with multi-GPU support. The unified interface ensures that the extraction logic remains independent of the backend provider.

\noindent\textbf{\textit{2-Pass Reflection.}} Quality filtering employs 2-pass reflection: a first pass extracts candidate entities with confidence scores, and a second pass scores them on relevance, accuracy, and consistency (0–1 each). Entities below configurable thresholds (default: 0.7) are filtered out, at the cost of one extra LLM call per chunk but with fewer false positives.
% 这个 adding这里没看懂？啥意思？adding干啥用的？

\subsection{Interactive Analysis Dashboard}
The \textit{VISUALIZATION} module is built with Next.js, React, and TypeScript; the dashboard features 5 tabbed visualization modes.

\noindent\textbf{\textit{Visualization Components.}} Interactive Map uses Mapbox GL for heatmap clusters; Multi-Track Timeline employs D3.js for category-based events; Entity Network renders ReactFlow co-occurrence graphs; Basic Timeline lists temporal entities chronologically; Dimension Breakdown shows frequency distributions.

\noindent\textbf{\textit{Spatiotemporal Analytics.}} Built-in spatiotemporal algorithms include DBSCAN-inspired clustering with 50km spatial and 7-day temporal radii, sliding-window burst detection, and co-occurrence network analysis. Four analytics panels provide quality metrics, burst detection, temporal analytics, and spatial visualization.

\begin{figure}[t]
	\centering
{\includegraphics[width=3.3 in]{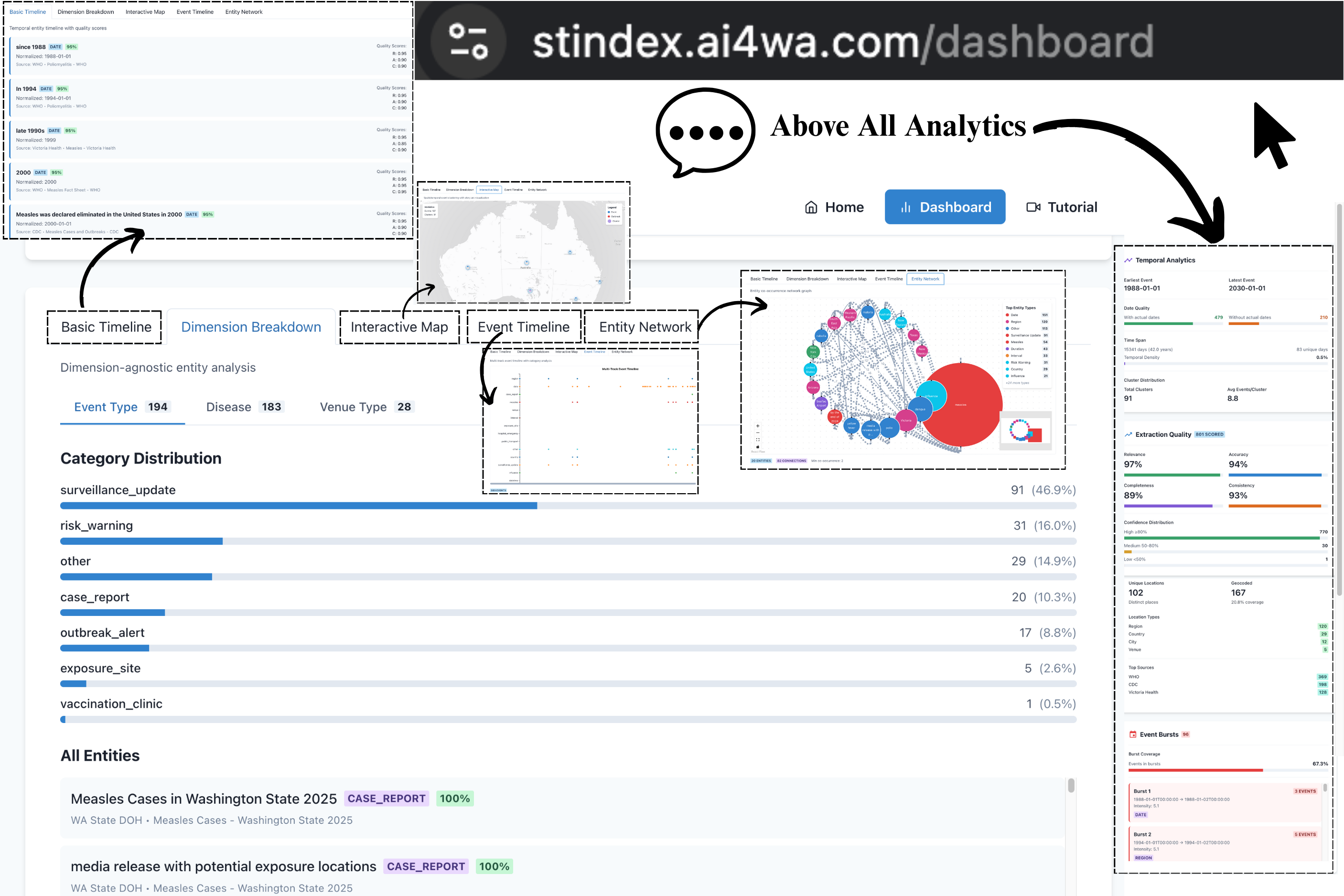}\label{fig:temporal_analytics}}%
	\hfil
 %    \captionsetup[subfloat]{labelfont=scriptsize,textfont=scriptsize}
	% \subfloat[Spatial Analytics]{\includegraphics[width=1.7in,height=1.3in]{Figure/spatial_analytics.png}\label{fig:spatial_analytics}}%
	% \hfil
 % 	\subfloat[Interactive Map]{\includegraphics[width=1.7in,height=1.3in]{Figure/map.png}\label{fig:map}}%
	% \hfil
 % 	\subfloat[Entity Network]{\includegraphics[width=1.6in,height=1.3in]{Figure/network.png}\label{fig:network}}%
	% \hfil

    \caption{STIndex Dashboard Demonstration}
    \Description{Four screenshots of the STIndex dashboard}
    \label{fig:dashboard}
\end{figure}

\section{Demonstration}

\subsection{Case Study: Public Health Surveillance}
We demonstrate STIndex on public health surveillance using Qwen3-8B, processing 10 real-world documents with 115 chunks. The system extracts 801 entities across five dimensions: 229 temporal, 167 spatial, 183 disease, 194 event types, and 28 venue types. Context-aware extraction preserves inter-dimensional relationships within single calls—each measles exposure event maintains linked temporal, spatial, event type, venue, and disease attributes. DBSCAN clustering generates 91 spatiotemporal clusters with 2--4 events each (mean: 2.3), successfully grouping related exposures while distinguishing outbreak waves. Figure~\ref{fig:dashboard} showcases the analytical and visualisation components. Details are available in the deployed dashboard at \url{https://stindex.ai4wa.com/dashboard}.

\begin{table}[t]
\centering
\caption{Overall Evaluation Results (500 Document Chunks)}
\label{tab:evaluation}
\small
\begin{tabular}{@{}llrrrr@{}}
\toprule
\textbf{Model} & \textbf{Mode} & \textbf{T-P} & \textbf{T-R} & \textbf{T-F1} & \textbf{Comb-F1} \\
\midrule
\multirow{2}{*}{GPT-4o-mini\cite{gpt4omini2024}} & Baseline & 67.83 & 65.44 & 66.61 & 70.72 \\
& \textbf{STIndex} & \textbf{71.05} & \textbf{68.32} & \textbf{69.66} & \textbf{73.81} \\
\cmidrule{2-6}
& \textit{Improvement} & \textit{+3.22} & \textit{+2.88} & \textit{+3.05} & \textit{+4.37\%} \\
\midrule
\multirow{2}{*}{Qwen3-8B\cite{qwen32025}} & Baseline & 59.80 & 56.82 & 58.27 & 69.81 \\
& \textbf{STIndex} & \textbf{67.84} & \textbf{65.22} & \textbf{66.50} & \textbf{72.32} \\
\cmidrule{2-6}
& \textit{Improvement} & \textit{+8.04} & \textit{+8.40} & \textit{+8.23} & \textit{+3.60\%} \\
\midrule
\multicolumn{6}{c}{} \\
\toprule
\textbf{Model} & \textbf{Mode} & \textbf{S-P} & \textbf{S-R} & \textbf{S-F1} & \textbf{MDE (km)} \\
\midrule
\multirow{2}{*}{GPT-4o-mini\cite{gpt4omini2024}} & Baseline & 87.11 & 65.59 & 74.83 & 377.32 \\
& \textbf{STIndex} & \textbf{92.00} & \textbf{67.65} & \textbf{77.97} & \textbf{369.02} \\
\cmidrule{2-6}
& \textit{Improvement} & \textit{+4.89} & \textit{+2.06} & \textit{+3.14} & \textit{+2.2\%} \\
\midrule
\multirow{2}{*}{Qwen3-8B\cite{qwen32025}} & Baseline & 89.40 & 74.63 & \textbf{81.35} & 1371.88 \\
& \textbf{STIndex} & \textbf{88.39} & \textbf{70.03} & 78.15 & \textbf{444.15} \\
\cmidrule{2-6}
& \textit{Improvement} & \textit{-1.01} & \textit{-4.60} & \textit{-3.20} & \textit{+67.6\%} \\
\bottomrule
\multicolumn{6}{l}{\scriptsize T-P/R/F1: Temporal Precision/Recall/F1 (\%), S-P/R/F1: Spatial Precision/Recall/F1 (\%)} \\
\multicolumn{6}{l}{\scriptsize Comb-F1: Combined F1 (\%), MDE: Mean Distance Error (km)} \\
\end{tabular}
\end{table}
% \vspace{-4cm}
\subsection{Evaluation}

  \noindent\textbf{\textit{Setup.}} We evaluate STIndex against a baseline approach that processes each document chunk
  independently without context. The baseline extracts spatiotemporal entities from each chunk in isolation, while
  STIndex maintains extraction context across chunks within the same document, enabling resolution of relative temporal
  expressions and disambiguation of location mentions. We use a synthetic dataset of 500 document chunks annotated with ground
   truth spatiotemporal entities, generated by Claude Sonnet 4.5 \cite{claudecode2024} and followed by human review. We group chunks into 6 categories, from straightforward/baseline to standard cases. Harder types involve overlapping temporal/spatial references, relative time expressions, and underspecified locations.
   We compare GPT-4o-mini vs.\
  Qwen3-8B in baseline vs.\ STIndex mode. 
  % 这个value exact mode, comparing, 这句话不通吧？
  Temporal values normalized to ISO 8601 and evaluated by exact match. Spatial matching evaluated by fuzzy match with a substring/word overlap threshold of $\geq$50\%. % 这里的这个with用if? 和上面感觉一样的问题，没说清楚
We compute precision, recall, and F1 scores for each dimension, plus normalization accuracy for temporal
  entities, geocoding success rate, and mean distance error (MDE) for spatial entities.
  % normalization, 还有另外几个错误，没有定义，至少要解释下

\noindent\textbf{\textit{Results.}} Table~\ref{tab:evaluation} presents results across 100 documents. STIndex
improves combined F1 by 4.37\% for GPT-4o-mini and 3.60\% for Qwen3-8B over baseline. GPT-4o-mini shows
temporal precision improvements (+3.22 percentage points) and spatial precision gains (+4.89 pp), while Qwen3-8B
achieves larger temporal F1 gains (+8.23 pp) from improved recall. Notably, STIndex improves spatial
geocoding accuracy, reducing MDE by 67.6\% for Qwen3-8B (1372km $\rightarrow$ 444km) compared to %
GPT-4o-mini's 2.2\% reduction, demonstrating STIndex's context-aware extraction better handles ambiguous location
names despite extracting fewer spatial entities. GPT-4o-mini achieves higher precision (temporal: 71.05\% vs.\ %
67.84\%, spatial: 92.00\% vs.\ 88.39\%) and combined F1 (73.81\% vs.\ 72.32\%). However, Qwen3-8B shows superior geocoding accuracy, with a 67.6\% MDE reduction compared to GPT-4o-mini’s 2.2\%.

% Both models achieve perfect
% temporal normalization accuracy (100\%) and geocoding success (100\%), validating post-processing tool effectiveness.

\section{Conclusion}

STIndex addresses fragmentation in spatiotemporal information extraction by unifying preprocessing, extraction, and visualization in an end-to-end system. Innovations include configurable dimensions for domain-agnostic extraction, context-aware processing with document-level memory for resolving ambiguous references, and client-side dashboard implementation for visualization and analytics. We demonstrate and validate the system through a real-world case study and a systematic evaluation of synthetic documents. Future work will integrate backend APIs for real-time extraction, enable cross-component filtering for synchronized exploration across views, and add export capabilities for extracted data.


\bibliographystyle{ACM-Reference-Format}
\bibliography{ref}

\end{document}